**Title**

Biomechanical analysis based on a full-body musculoskeletal model for evaluating the effect of a passive lower-limb exoskeleton on lumbar load

**Author affiliation**

Naoto Haraguchi [a,*], Kazunori Hase [a]

a. Department of Mechanical Systems Engineering, Tokyo Metropolitan University, 1-1 Minami-Osawa, Hachioji-city, Tokyo 192-0397, Japan

Corresponding author: Naoto Haraguchi

Department of Mechanical Systems Engineering, Tokyo Metropolitan University, 1-1 Minami-Osawa, Hachioji-city, Tokyo 192-0397, Japan

Telephone number: +81-42-677-1801

E-mail address: haraguchi-naoto1@ed.tmu.ac.jp

**Abstract**

The present study investigated the effect of a passive lower-limb exoskeleton on lumbar load and verified the effectiveness of biomechanical analysis for evaluating the physical burden while wearing the exoskeleton. Twelve healthy male participants performed an assembly task under three conditions: standing and high and low sitting while wearing the exoskeleton. We mainly analyzed the joint compression force computed using the musculoskeletal model based on measurements obtained from a motion capture system and force platform. While wearing the exoskeleton, the lumbar joint compression force increased. Analysis of the joint compression force can determine the increase in the lumbar load when wearing an exoskeleton, which is not possible using the traditional analysis of posture and muscular activity. Therefore, biomechanical analysis based on a full-body musculoskeletal model provides valuable information in evaluating the effect of the exoskeleton that attempts to prevent musculoskeletal disorders.

**Practitioner Summary**

A passive lower-limb exoskeleton helps to prevent musculoskeletal disorders. However, investigation of the effect of the exoskeleton on the lumbar burden is rare. We herein clarify the effect of the exoskeleton on the lumbar burden and provide a method by which to evaluate the physical burden while wearing the exoskeleton.

**Main text**

**1. Introduction**

The World Health Organization has mentioned the necessity of preventing musculoskeletal disorders, such as lower back pain and osteoarthritis (World Health Organization, 2018). One cause of musculoskeletal disorders is the physical burden while working (Driscoll et al., 2014; Richmond et al., 2013). Devices that reduce the physical burden during labor have been developed to prevent musculoskeletal disorders. Furthermore, the effect of the devices on the physical burden have been quantitatively analyzed (Iranzo et al., 2020; Koopman et al., 2020). Analysis of the devices are critical to enable the development and improvement of devices for preventing musculoskeletal disorders.

The present study focused on a passive lower-limb exoskeleton (Figure 1). The exoskeleton attached to the lower limb supports the weight of the wearer and reduces the burden on the lower limbs. The muscle burden of the lower limbs has been reported to have been reduced by wearing the exoskeleton, which effectively prevents musculoskeletal disorders of the lower limbs (Yan et al., 2021; Kim et al., 2013; Hasegawa and Muramatsu, 2013). In contrast, a previous study reported that the burden on the lower back may increase due to changes in posture and muscle activity when wearing the exoskeleton (Luger et al., 2019b). Since the burden on the lower back may lead to the risk of developing lower back pain, accurately evaluating the burden on the lower back is necessary. However, the traditional analysis of posture and muscular activity has not accurately evaluated the lumbar burden when wearing an exoskeleton (Luger et al., 2019b). Previous studies have only analyzed muscular activity to evaluate the effect of the exoskeleton on the physical burden (Yan et al., 2021; Kim et al., 2013; Hasegawa and Muramatsu, 2013). Therefore, evaluation of the lumbar burden using the existing evaluation method seems complicated. To clarify the effect of the exoskeleton on the lumbar burden, establishing a new evaluation method is necessary.

The lumbar burden was evaluated in previous studies by analyzing the joint load (Andersson et al., 1980; Mcgill and Norman, 1985). A method to perform these analyses by musculoskeletal models has been applied to evaluations of lumbar burden from physical support devices (Imamura et al., 2011). Therefore, analyzing the joint load to evaluate the lumbar burden when wearing an exoskeleton is practical, but the external forces acting on the body need to be measured using musculoskeletal models. There are two external forces when an exoskeleton is used: the ground reaction force acting on the foot and the exoskeleton acting on the lower limbs. Therefore, all external forces acting on the body for musculoskeletal model analysis when wearing the exoskeleton need to be measured.

The present study attempts to evaluate the effect of the exoskeleton on the lumbar burden by joint load analysis. For this purpose, we analyzed the lumbar joint load using musculoskeletal model analysis based on measuring physical movement and external forces when wearing the exoskeleton. In addition, we propose more ergonomic use and improvement of the exoskeleton to reduce the lumbar burden. Furthermore, we analyzed the burden on the lower back, lower limbs, and neck based on biomechanical analysis to verify the effectiveness of the proposed method for evaluating the effects of the exoskeleton on the physical burden.

## 2. Materials and methods

### 2.1. Participants

The participants were 12 healthy males (height: 1.75 ± 0.06 m, weight: 65.0 ± 8.0 kg, age: 24.3 ± 3.1 years). The present study was approved by the Ethics Committee of Tokyo Metropolitan University. All participants received verbal and written explanations of the study content before the experiment and gave their written informed consent.

## 2.2. Conditions

The exoskeleton shown in Figure 1 (ChairlessChair®; noonee AG, Switzerland) was used in the experiment. This product is widely used in manual labor occupations and consists of an exoskeleton that supports the body weight and straps that enable the device to be worn. The exoskeleton part is composed of the seat, the upper link, and the lower link. The bottom of the lower link touches the ground. The seat, upper link, and lower link are jointed by one degree of freedom. The joint between the upper and lower links is constrained by rotation at any pre-set angle.

The experimental condition was set referring to a previous study (Luger et al., 2019b). We consider three experimental conditions: standing without the exoskeleton (*STAND*) and high and low sitting with the exoskeleton (*HIGH_SEAT* and *LOW_SEAT*). In the *HIGH_SEAT* and *LOW_SEAT* states, the knee joint flexion angles when wearing the exoskeleton are set to 125° and 90°, respectively. The participants were not instructed on other postures, and each participant remained in a relaxed posture in each experimental condition. The working height (*WH*) and the working distance (*WD*) are determined, respectively, by the following equations (Luger et al., 2019a):

$$WH = height_{elbow-floor} - (\sin(\alpha) \cdot length_{lower\ arm}) \quad (1)$$

$$WD = length_{lower\ arm} + (\sin(\beta) \cdot length_{upper\ arm}) \quad (2)$$

where $height_{elbow\text{-}floor}$ is the height from the floor to the olecranon in the elbow, $length_{lower\ arm}$ is the length from the olecranon in the elbow to the triquetrum in the hand, $length_{upper\ arm}$ is the length from the olecranon in the elbow to the acromion in the shoulder, α is the angle (15°) of the lower arm below the horizontal, and β is the angle (20°) of the shoulder in ante-version.

## 2.3. Procedure

Participants were habituated to wearing the exoskeleton and performing the experimental task, which mimicked assembly work in a factory. Participants stood in front of a work table with the exoskeleton attached (or not attached) and were asked to assemble two plates with bolts and nuts using a ratchet wrench, which was used by the dominant hand. The working time was set to 80 s. The working pace was set to complete the assembly of one bolt and nut in 30 s, which was determined to be achievable for all participants. Three trials were measured under each condition, and a three-minute break was set between each measurement. The measurement sequence was randomized to prevent an order effect.

## 2.4. Measurements

### 2.4.1. Physical motion

The motion capture system (OptiTrack Flex3; Natural Point Inc., USA) measured the three-dimensional coordinates of markers attached to the whole body. A total of 77 reflection markers were placed on the body referring to the marker arrangement of the musculoskeletal model (Raabe and Chaudhari, 2016), and five locations were added for joint angle calculation. Data were digitally filtered (low-pass filter, fourth order, -3 dB at 6 Hz, Butterworth) and were sampled at 100 Hz.

### 2.4.2. External force

The ground reaction force acting on the foot was measured by the mobile force plate system (M3D-FP-U; Tec Gihan Co., Ltd., Japan) shown in Figure 2. The mobile force plate system measures the ground reaction force acting on the foot (excluding the force acting on the exoskeleton) by attaching a six-axis force plate with an amplifier and an AD converter to the foot. Data were digitally filtered (low-pass filter, fourth order, -3 dB at 18 Hz, Butterworth) and were sampled at 1,000 Hz. There are two methods to measure the force generated by the exoskeleton: a method to measure the force acting on the exoskeleton from the floor and a method to measure the force acting on the thigh from the seat of the exoskeleton. In the latter method, measurement data can be input directly into the musculoskeletal model. Therefore, the force acting on the thigh attached to the exoskeleton was measured by the six-axis force sensor (USX10-H10-500N; Tec Gihan Co., Ltd., Japan) shown in Figure 2. The six-axis force sensors were fixed to the seat of the exoskeleton via a bracket. Data were recorded using a data logger with a built-in amplifier and an AD converter (GDA-12B; Tec Gihan Co., Ltd., Japan) and were digitally filtered (low-pass filter, fourth order, -3 dB at 18 Hz, Butterworth) and sampled at 1,000 Hz. In addition, the force plates and force sensors are equipped with brackets attached to reflection markers. The position and posture of each sensor coordinate can be calculated from the marker coordinates measured by the motion capture system.

### 2.4.3. Muscular activity

Activities of the upper trapezius, erector spinae, and medial head of the gastrocnemius were obtained by surface electromyography (EMG) recording surface electrodes with a built-in amplifier (DL-142; S&ME Inc., Japan). A ground earth electrode was placed over the styloid process of the ulna. Data were recorded by the measurement system equipped with an AD converter (DSS300-U; Tec Gihan Co., Ltd., Japan) and were sampled at 5,000 Hz. Recorded data were digitally filtered (high-pass filter, 11th order, -3 dB at 20 Hz, Butterworth), and the root means square (RMS; 250-ms moving window with 50% overlap) was calculated. Before the experiment, the EMG was collected during 20-s submaximal reference voluntary contractions (Luger et al., 2019b). Trapezius: the participant stood straight with his feet hip-width apart, kept both arms in 90° abduction, and held 1 kg in each hand. Erector spinae: the participant was lying prone with his lower body on a bench that extended until reaching his hips, keeping his upper body horizontal. Gastrocnemius: the participant stood in front of the table while bending his lower back 90°, and placed his lower arm on the table. A 10-kg weight was attached to the lower back, and the participant had to keep his ankles in 45° plantar flexion.

## 2.5. Data analysis

### 2.5.1. Ground reaction force and exoskeleton reaction force

After converting the coordinates of the measured values of the mobile force plate system and the six-axis force sensor based on the three-dimensional coordinates of the reflective markers, the force vector of the ground reaction force acting on the foot and the exoskeleton reaction force acting on the thigh were calculated. We calculated the norm of the force vector, which is normalized by the body weight and gravitational acceleration, i.e., % body weight, of each participant.

### 2.5.2. Joint angle, joint torque, and joint compression force

The joint angle was calculated by the three-dimensional coordinates of the reflective markers for comparison with a previous study (Luger et al., 2019b) and was analyzed in detail. Three angles were calculated in the sagittal plane between the upper trunk and the lower trunk (trunk flexion angle), the upper trunk and the vertical axis (trunk inclination angle), and the neck and the vertical axis (neck inclination angle). As shown in Figure 3, the upper trunk, the lower trunk, and the neck were defined as lines connecting the reflective markers at C7, T3, T10, L5, and the tragus.

The joint torque and joint compression force in the lower back and lower limbs were analyzed using the standard, open-source musculoskeletal model analysis software, OpenSim 4.1 (Delp et al., 2007). We used a previously developed musculoskeletal full-body model (Raabe and Chaudhari, 2016). The musculoskeletal model consists of 21 segments, 30 degrees of freedom, and 324 muscles. The ankle and knee joints have one degree of freedom in dorsiflexion-plantarflexion and flexion-extension, respectively. The hip and lumbar (L4L5) joints have three degrees of freedom. The pelvic segment has a virtual joint of six degrees of freedom between the spatial coordinates. The residual force due to measurement error is generated in the virtual joint. The analysis procedure followed a series of standard steps in OpenSim, as shown in Figure 4. First, the joint angle is calculated by inverse kinematics based on the measured data of the motion capture system. The joint torque is calculated by inverse dynamics based on the joint angle and measured data of external forces. After computing the muscle force by static optimization based on the joint torque, joint angle, and external forces, the joint reaction force is calculated by the muscle force and external forces. The joint compression force is defined as the component of the joint reaction force acting in the axial direction. We analyzed the joint torque in the sagittal plane and the joint compression force at the L4L5, hip, knee, and ankle dominant joints. Before analyzing the joint torque and joint compression force, the effect of measurement error was reduced by the residual reduction algorithm, which regulates the parameters of the musculoskeletal models (Delp et al., 2007).

Since the musculoskeletal model has a unified head and trunk and no neck joint is defined, the neck joint torque generated by the head weight was calculated directly by the reflective markers. The neck joint was defined as the midpoint between the marker at C7 and the clavicle (Syamala et al., 2018). The mass center of the head was estimated as 17% of the distance from the mid-tragus (midpoint of the left and right tragus markers) to the top of the head marker (Vasavada et al., 2015). The head weight was defined as follows, using an estimation based on head circumference and weight (Clauser et al., 1969):

$$mass_{head} = 0.104 \cdot circumference_{head} + 0.015 \cdot mass_{body} - 2.189 \quad (3)$$

where $mass_{head}$ is the head mass [kg], $circumference_{head}$ is the head circumference around the glabella [cm], and $mass_{body}$ is the body weight [kg].

### 2.5.3. Electromyography

The RMS of the experimental recordings was normalized to the RMS of the most stable 5-s period of each RVC, i.e., reference voluntary electrical activity (%RVE; Mathiassen et al., 1995). We calculated the RMS of each muscle on the dominant side for each participant.

### 2.6. Statistical analysis

Statistical analysis was performed on the average of the data for all measurement times. The exoskeleton reaction force was compared between the *HIGH_SEAT* and *LOW_SEAT* conditions. We checked for a normal distribution based on the Shapiro-Wilk test and performed a paired t-test. The ground reaction force, joint angle, joint torque, joint compression force, and EMG were compared among the *STAND*, *HIGH_SEAT*, and *LOW_SEAT* conditions. We performed Friedman's test on the ankle joint compression force, which was confirmed to have a non-normal distribution using the Shapiro-Wilk test, and the significant main effects were further explored using Holm's SRB procedure. We performed a one-factor repeated-measures analysis of variance (RM-ANOVA) on the other parameters that were confirmed to have a normal distribution by the Shapiro-Wilk test, and the significant main effects were further explored using Shaffer's MSRB procedure. All statistical analyses were performed with R Commander 2.7.0., Windows R4.0.2 (GRAN, freeware), and $p < 0.05$ indicated statistical significance.

## 3. Results

### 3.1. Ground reaction force and exoskeleton reaction force

The ground reaction force (Figure 5 (a)) was significantly affected by the experimental condition. The ground reaction force when wearing the exoskeleton was smaller compared to *STAND* ($p < 0.0001$), and that for *LOW_SEAT* was smaller compared to that for *HIGH_SEAT* ($p = 0.0179$). Compared to the value in *STAND* (101.4% body weight (±3.7% body weight)), that in *HIGH_SEAT* was decreased by 71% (29.4% body weight (±5.8% body weight)), and that in *LOW_SEAT* was decreased by 75% (24.9% body weight (±6.1% body weight)).

The exoskeleton reaction force (Figure 5 (b)) was significantly different between these two conditions ($p = 0.0006$), and the exoskeleton reaction force for *LOW_SEAT* (83.1% body weight (±8.1% body weight)) increased significantly compared to that for *HIGH_SEAT* (75.6% body weight (±8.7% body weight)).

## 3.2. Joint angle, joint torque, and joint compression force

The trunk inclination angle (Figure 6 (a)) was significantly affected by the experimental conditions. The trunk inclination angle when wearing the exoskeleton was larger than that for *STAND* ($p = 0.0054$). The neck inclination angle (Figure 6 (b)) and the trunk flexion angle (Figure 6 (c)) were not significantly affected by the experimental conditions.

The joint torques for L4L5 (Figure 7 (a)), hip (Figure 7 (b)), knee (Figure 7 (c)), and ankle (Figure 7 (d)) were significantly affected by the experimental conditions. The L4L5 joint extension torque when wearing the exoskeleton was larger than that for *STAND* ($p = 0.0004$). The L4L5 joint extension torque in *HIGH_SEAT* was 31.1 N·m (±11.2 N·m) and in *LOW_SEAT* was 33.8 N·m (±15.3 N·m). The hip joint flexion torque when wearing the exoskeleton was smaller than that for *STAND* ($p = 0.0002$). The knee joint torque was significantly different and the flexion-extension direction changed between the experimental condition wearing the exoskeleton and *STAND* ($p = 0.0008$). The ankle joint plantarflexion torque when wearing the exoskeleton was smaller than that for *STAND* ($p < 0.0001$), and that for *LOW_SEAT* was smaller compared to that for *HIGH_SEAT* ($p = 0.0382$). The neck joint torque (Figure 7 (e)) was not significantly different under the experimental conditions.

The joint compression forces for L4L5 (Figure 8 (a)), hip (Figure 8 (b)), knee (Figure 8 (c)), and ankle (Figure 8 (d)) were significantly affected by the experimental conditions. The L4L5 joint compression force when wearing the exoskeleton was larger than that for *STAND* ($p = 0.0006$). The L4L5 joint compression force for *HIGH_SEAT* was 1,253 N (±324 N) and that for *LOW_SEAT* was 1,296 N (±401 N). The hip joint compression force when wearing the exoskeleton was smaller than that for *STAND* ($p = 0.0008$). Compared to the value for *STAND* (859 N (±296 N)), that for *HIGH_SEAT* was decreased by 50% (429 N (±71 N)) and that for *LOW_SEAT* was decreased by 53% (401 N (±87 N)). The knee joint compression force when wearing the exoskeleton was smaller than that for *STAND* ($p < 0.0001$), and that for *LOW_SEAT* was smaller compared to that for *HIGH_SEAT* ($p = 0.0001$). Compared to the value for *STAND* (748 N (±216 N)), that for *HIGH_SEAT* was decreased by 80% (147 N (±35 N)), and that for *LOW_SEAT* was decreased by 87% (97 N (±66 N)). The ankle joint compression force when wearing the exoskeleton was smaller than that for *STAND* ($p = 0.0015$), and that for *LOW_SEAT* was smaller than that for *HIGH_SEAT* ($p = 0.0034$). Compared to the value for *STAND* (823 N (±226 N)), that for *HIGH_SEAT* was decreased by 69% (255 N (±83 N)), and that for *LOW_SEAT* was decreased by 75% (206 N (±92 N)).

### 3.3. Electromyography

Only EMG of the gastrocnemius was significantly affected by the experimental conditions (Figure 9); EMG of the gastrocnemius (Figure 9 (c)) when wearing the exoskeleton was smaller than that for *STAND* ($p = 0.0081$).

## 4. Discussion

The primary goals of the present study were to evaluate the effect of the exoskeleton on lumbar load, and to verify the effectiveness of the evaluation method. First, we verified the validity of the measurement of the external force and the computation of the joint load. Next, we evaluated the lumbar load when wearing the exoskeleton based on the joint load analysis. Finally, we verified the effectiveness of the evaluation method compared to the conventional method.

### 4.1. Validation of the analysis

The ground reaction force decreased when wearing the exoskeleton in the present study, as reported in previous studies (Luger et al., 2019b; Hasegawa and Muramatsu, 2013). In addition, the ground reaction force decreased for *LOW_SEAT* compared with *HIGH_SEAT*, and the exoskeleton reaction force increased for *LOW_SEAT* compared with *HIGH_SEAT*. These variations are attributed to the weight transfer to the exoskeleton. Furthermore, since the motion in this experiment is nearly static, the total external force acting on the body is assumed to be equal to the body weight. The sum of the ground reaction force and the exoskeleton reaction force is approximately equal to the body weight. Therefore, all external forces acting on the body when wearing the exoskeleton could be measured in this experiment.

The L4L5 joint compression force increased when wearing the exoskeleton. This increase is attributed to increasing the L4L5 joint torque and the trunk inclination angle, as shown in Figure 10. The wearer must keep the center of gravity ahead of the base of support to prevent from falling backward. Therefore, the wearer moves the center of gravity forward by trunk inclination. This movement increases the moment arm of torque generated in the L4L5 joint by the upper body weight, and the L4L5 joint extension torque is increased. Thus, the muscle force of the lumbar extension muscle group exerting the L4L5 joint extension torque is increased. Lumbar joints are compressed further by the increase in muscle force of the lumbar extension muscle group, which led to an increase in the L4L5 joint compression force. A previous study reported that the L4L5 joint load increases when sitting in a chair as compared to standing (Callaghan and McGill, 2001). In addition, it has been reported that the L4L5 joint torque in slouched seating is 32.3 N·m, and the L4L5 joint compression force in slouched seating is 1,181 N (Castanharo et al., 2014). These reports are similar to the results of the present study, which supports the validity of the L4L5 joint load analysis.

When wearing the exoskeleton, the joint compression force in the ankle and knee decreased, which is attributed to the decrease in the ground reaction force. The ankle joint plantarflexion torque was also decreased when wearing the exoskeleton, and the knee joint torque acted in the extension direction. Thus, the force of the gastrocnemius muscle contributing to the plantarflex in the ankle and the flex in the knee is assumed to decrease when wearing the exoskeleton. The present study and a previous study confirmed a decrease in muscular activity in the gastrocnemius when wearing the exoskeleton (Luger et al., 2019b), which supports the validity of the knee and ankle joint analysis.

The hip joint compression force decreased when wearing the exoskeleton. Since the hip joint is affected by the ground reaction force and the exoskeleton reaction force, the joint load by external forces does not change when standing or wearing the exoskeleton. However, the hip joint flexion torque is decreased by sitting on the exoskeleton attached to the thigh near the buttocks. The decrease in the hip joint flexion torque leads to a decrease in the muscle force of the hip joint flexion muscle group, which leads to a decrease in the hip joint compression force. A previous study reported that the hip joint torque and joint compression force are decreased during sitting (Bergmann et al., 2001), which supports the validity of the hip joint load analysis.

There was no difference in the neck joint torque under the experimental conditions. The invariance is due to the lack of difference in the moment arm of the neck joint torque generated by the head weight because there was no difference in the neck inclination angle. Furthermore, there is no difference in muscular activity in the upper trapezius, which supports the validity of the neck joint load analysis.

**4.2. Evaluation of the lumbar load**

The increase in the L4L5 joint compression force when wearing the exoskeleton resulted in a greater lumbar burden. Considering that the L4L5 joint load when wearing the exoskeleton was similar to that when slouched sitting in a chair, the lumbar burden when wearing the exoskeleton is assumed to be similar to the lumbar burden when slouched sitting in a chair. Therefore, it is unlikely that simply wearing the exoskeleton significantly increases the lumbar burden or the risk of developing lower back pain. However, prolonged use of an exoskeleton that exerts a similar burden while sitting in a chair is not preferable because prolonged sitting places strain on the lumbar region. Therefore, an appropriate standing posture when wearing the exoskeleton is recommended to reduce the lumbar burden.

Since the increase in lumbar burden when wearing the exoskeleton is due to the trunk inclination to prevent falling backward, the lumbar burden can be reduced by installing a function to prevent falling backward in the exoskeleton. For example, the wearer can maintain an upright posture by moving the part of the exoskeleton touching the ground backward, farther away from the body. The upright posture not only suppresses an increase in the lumbar joint torque by the trunk forward inclination, but also mitigates the increase in the lumbar burden when wearing the exoskeleton.

### 4.3. Effectiveness of the evaluation method

There was no significant difference in the muscular activity of the erector spinae under the experimental conditions, which did not match the increase in L4L5 joint load when wearing the exoskeleton, whereas the flexion relaxation phenomenon, for which muscular activity of the erector spinae disappears in the static flexive posture of the trunk and shifts to passive posture retention function, has been reported (Schultz et al., 1985). Thus, the trunk flexion angle affects the muscular activity of the erector spinae in the static posture when wearing the exoskeleton. There was no significant difference in muscular activity of the erector spinae and the trunk flexion angle in the present study, and a previous study reported that the muscular activity of the erector spinae decreases and the trunk flexion angle increases when wearing the exoskeleton (Luger et al., 2019b). Therefore, it is likely that the flexion relaxation phenomenon caused the mismatch between the joint load analysis and the muscular activity analysis. In other words, the muscular activity analysis does not match the lumbar burden when wearing the exoskeleton. In contrast, since the lumbar burden can be evaluated by the joint load analysis based on the musculoskeletal model, the joint load analysis by the musculoskeletal model is adequate for the evaluation of the lumbar burden when wearing the exoskeleton.

A previous study reported that a decrease in the ground reaction force and the muscular activity of the gastrocnemius when wearing the exoskeleton prevents disease, such as Achilles tendinitis (Luger et al., 2019b). On the other hand, osteoarthritis caused by joint load accounts for a large number of musculoskeletal diseases in workers (World Health Organization., 2018). Therefore, joint load analysis by the musculoskeletal model is adequate for evaluation of lower limb burden when wearing the exoskeleton with the goal of preventing musculoskeletal disease. However, considering that a decrease in the knee and ankle joint compression force is due to a decrease in the ground reaction force when wearing the exoskeleton, the joint load of the knee and ankle can be inferred from only the ground reaction force. In the present study, the rate of decrease in the ground reaction force is comparable to the rate of decrease in the knee and ankle joint compression force. In contrast, there is a difference between the rate of decrease in the hip joint compression force and the ground reaction force. Since the hip joint is affected by the reaction force from the exoskeleton, the decrease in the hip joint compression force does not match the decrease in the ground reaction force. Therefore, the musculoskeletal model analysis is effective for evaluation of the hip joint load when wearing the exoskeleton.

A previous study reported the correlation between the neck joint torque and the muscular activity of the upper trapezius (Schuldt and Harms-Ringdahl, 1988). In the present study, there was no difference in the neck joint torque or the muscular activity of the upper trapezius. Therefore, the neck joint load can be evaluated from the muscular activity analysis or the joint load analysis.

## 4.4. Limitations

The cost incurred by measuring the external forces is a disadvantage of the proposed method. For example, since the exoskeleton reaction force is measured by fixing the six-axis force sensor to the passive lower-limb exoskeleton, it is necessary to manufacture a component that fixes the sensor to the exoskeleton. However, if the burden on only the knee joint, ankle joint, and neck joint is evaluated, it can be evaluated by the muscular activity analysis and the ground reaction force analysis. Therefore, selecting an appropriate method according to the evaluation is essential.

The joint torque computed in the musculoskeletal model is generated by the muscle force excluding the passive resistance, such as ligaments. Thus, the L4L5 joint compression force is different from the actual force. When torque is exerted by a ligament close to the joint, a large force is required because the moment arm is short. Therefore, the actual L4L5 joint compression force generated when wearing the exoskeleton is larger than the calculation of the present study. To improve the accuracy of the calculations, a ligament needs to be added to the musculoskeletal model.

The ankle and knee joint compression forces decreased for *LOW_SEAT* compared to *HIGH_SEAT* due to the decrease in the ground reaction force. However, although the ankle joint torque decreased for *LOW_SEAT* compared to *HIGH_SEAT*, the muscular activity of the gastrocnemius did not change between the two conditions. In addition, a previous study reported that the muscular activity of the gastrocnemius decreased for *LOW_SEAT* compared to *HIGH_SEAT* (Luger et al., 2019b). Therefore, further investigation is needed to clarify the change in the lower limb burden between the *HIGH_SEAT* and *LOW_SEAT* conditions.

## 5. Conclusions

We evaluated the physical burden through musculoskeletal model analysis based on measuring physical movement and external forces when wearing an exoskeleton. We observed that the lumbar burden increases when wearing the exoskeleton due to the L4L5 joint compression force. Thus, we recommend an appropriate standing posture when wearing the exoskeleton to reduce the lumbar burden. In addition, the increase in the lumbar burden when wearing the exoskeleton can be reduced by moving the part of the exoskeleton touching the ground backward, farther away from the body, so the wearer can maintain an upright posture. Furthermore, since the musculoskeletal model analysis can be used to evaluate the lumbar burden and hip joint burden, which cannot be evaluated from the traditional method, and the burden of the whole body simultaneously, joint load analysis using the musculoskeletal model is effective as a method for evaluating the physical burden when wearing the exoskeleton.

  The present study has the disadvantage of an increase in the cost incurred when measuring the external forces. When the inverse dynamics are performed from the body end (head and hand) to the pelvis segment based on the recurrence calculation of the Newton-Euler method, the lumbar joint load can be calculated even if the external forces acting on the lower legs are not measured. Thus, only the lumbar burden can be evaluated without measuring the external forces acting on the lower legs. However, since the muscle force of the psoas major, which straddles the lower back and hip joint, cannot be calculated correctly, the effect of the psoas major on the lumbar joint compression force must be investigated. If this problem is solved, then the lumbar load when wearing the exoskeleton can be evaluated without measuring the external forces.

**Disclosure statements**

There is no potential conflict of interest regarding the present study.

for Osteoarthritis? A Systematic Review." *Journal of Orthopaedic & Sports Physical Therapy* 43 (8): 515–524. doi:10.2519/jospt.2013.4796.

Schuldt, K., and Harms-Ringdahl, K. 1988. "E.m.g./moment relationships in neck muscles during isometric cervical spine extension." *Clinical Biomechanics* 3 (2): 58–65. doi:10.1016/0268-0033(88)90046-0.

Schultz, A.B., Haderspeck-Grib, K., Sinkora, G., and Warwick, D.N. 1985. "Quantitative studies of the flexion-relaxation phenomenon in the back muscles." *Journal of Orthopaedic Research* 3 (2): 189–197. doi:10.1002/jor.1100030208.

Syamala, K.R., Ailneni, R.C., Kim, J.H., and Hwang, J. 2018. "Armrests and back support reduced biomechanical loading in the neck and upper extremities during mobile phone use." *Applied Ergonomics* 73: 48–54. doi:10.1016/j.apergo.2018.06.003.

Vasavada, A.N., Nevins, D.D., Monda, S.M., Hughes, E., and Lin, D.C. 2015. "Gravitational demand on the neck musculature during tablet computer use." *Ergonomics* 58 (6): 990–1004. doi:10.1080/00140139.2015.1005166.

World Health Organization. 2018. "Preventing Disease Through a Healthier and Safer Workplace."33

Yan, Z., Han, B., Du, Z., Huang, T., Bai, O., and Peng, A. 2021. "Development and testing of a wearable passive lower-limb support exoskeleton to support industrial workers." *Biocybernetics and Biomedical Engineering* 41 (1): 221–238. doi:10.1016/j.bbe.2020.12.010.

**Figure captions**

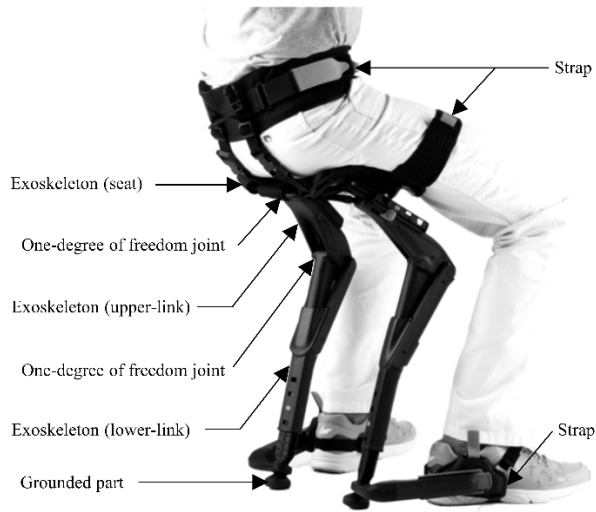

Figure 1. Passive lower-limb exoskeleton (ChairlessChair®). Photograph adapted from

https://www.noonee.com (©noonee AG).

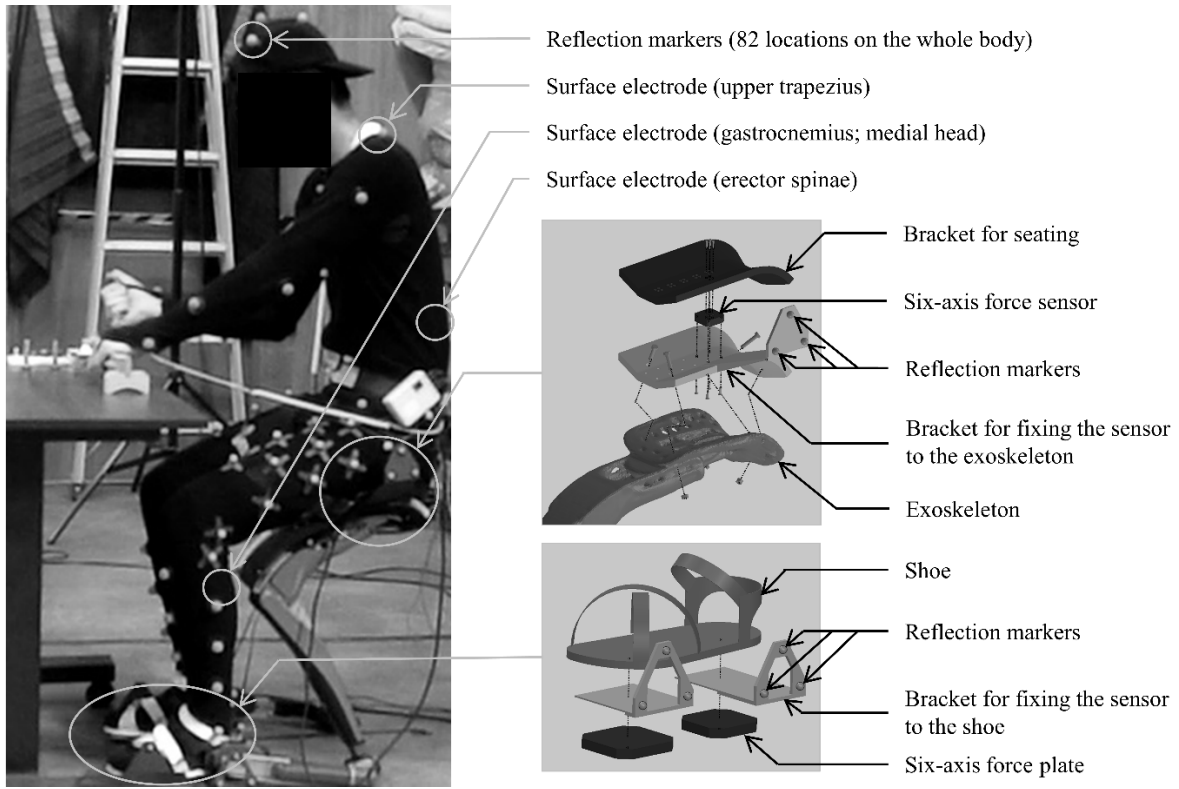

Figure 2. Overview of measurements. The reflection markers of the motion capture system are attached to 82 locations over the entire body. The surface electrodes are attached to the upper trapezius, the erector spinae, and the medial head of the gastrocnemius. The mobile force plate system measures the ground reaction force acting on the foot by the six-axis force plates attached to the bottoms of the shoes. The external force acting on the thigh is measured by the six-axis force sensor attached to the seat of the exoskeleton. The brackets attached to the reflection markers can measure the positions and postures of the force plates and sensors.

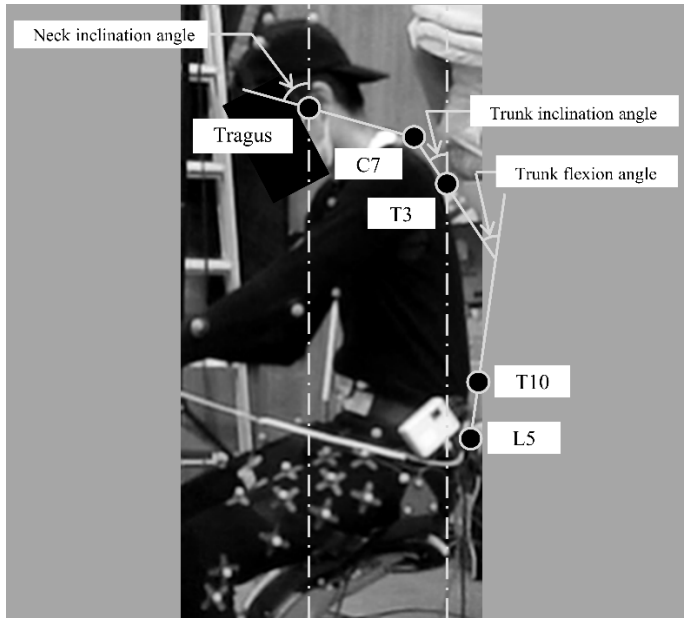

Figure 3. Joint angles. The trunk flexion angle was determined by the line passing through the L5 marker and T10 marker and the line passing through the T3 marker and C7 marker. The trunk inclination angle determined the line passing through the T3 marker and C7 marker and the vertical axis. The neck inclination angle was determined by the line passing through the C7 marker and the tragus marker and the vertical axis.

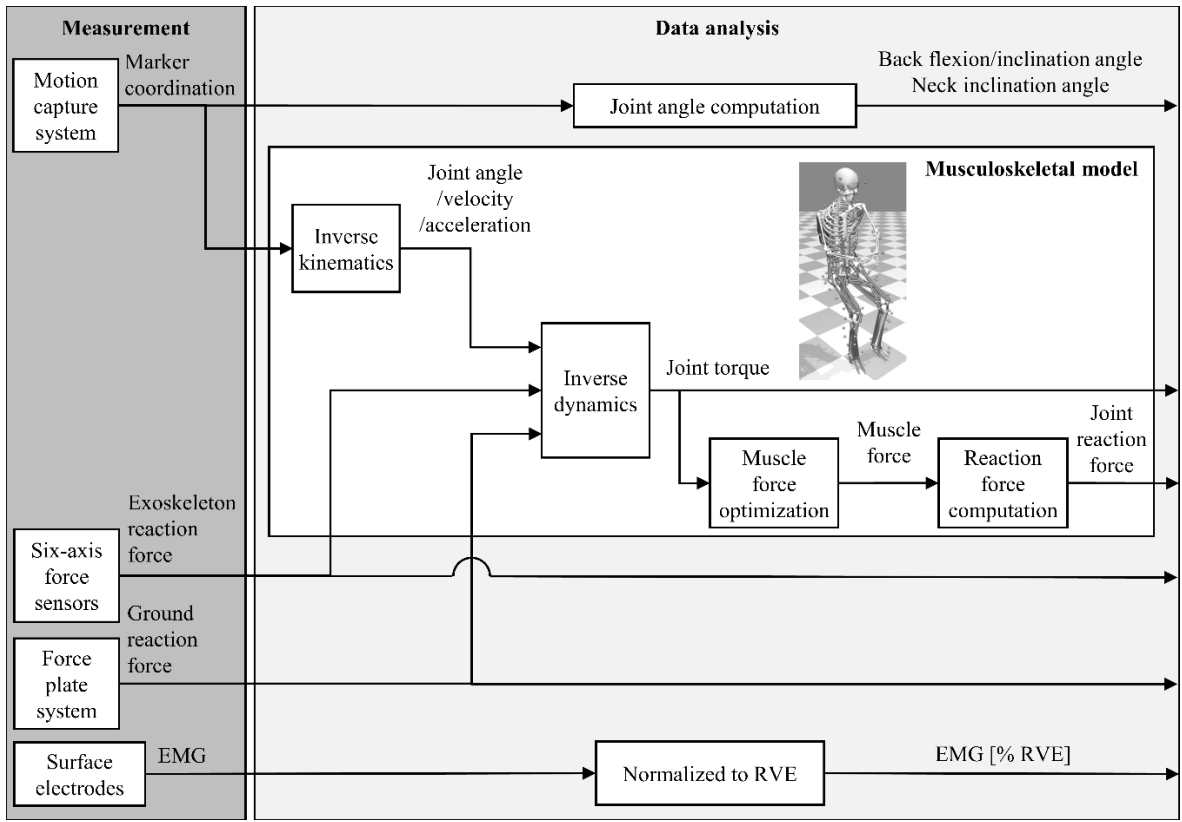

Figure 4. Overview of the data analysis. The joint angles of the trunk and neck are calculated by the measurement data of the motion capture system. Joint torque and reaction forces are calculated by the analysis using a musculoskeletal model. Analysis was performed using a musculoskeletal model to obtain the inverse kinematics and inverse dynamics and realize muscle force optimization and reaction force computation. The ground reaction force acting on the foot was calculated by the measurement data of the force plate system. The exoskeleton reaction force

acting on the thigh is calculated by the measurement data of the six-axis force sensors. The RMS

value is calculated by the EMG data and normalized to % RVE.

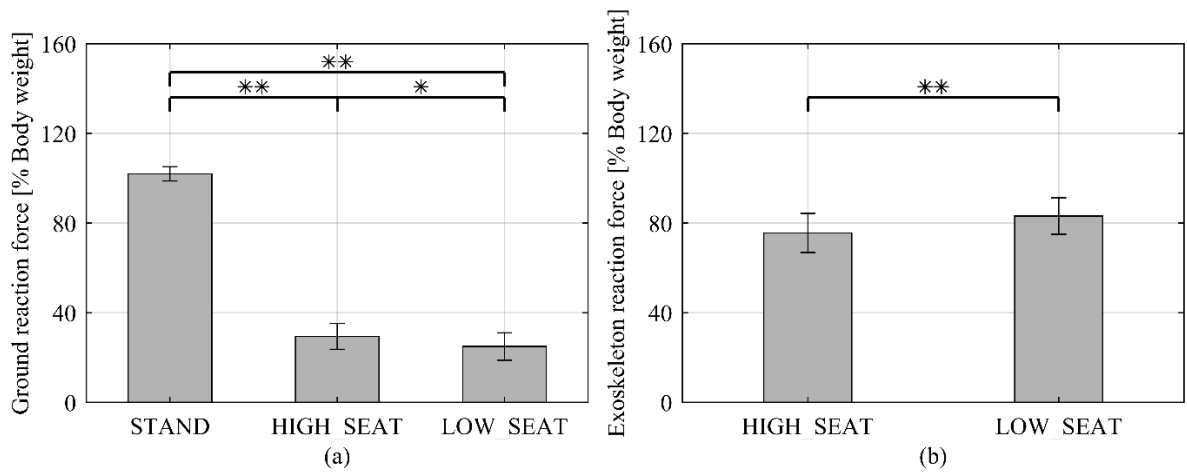

Figure 5. (a) Ground reaction force and (b) exoskeleton reaction force. *STAND* is the condition of standing without the exoskeleton. *HIGH_SEAT* and *LOW_SEAT* are conditions such that the knee joint flexion angles when wearing the exoskeleton are set to 125° and 90°, respectively. The significant differences between the conditions are indicated by * ($p < 0.05$) and ** ($p < 0.01$).

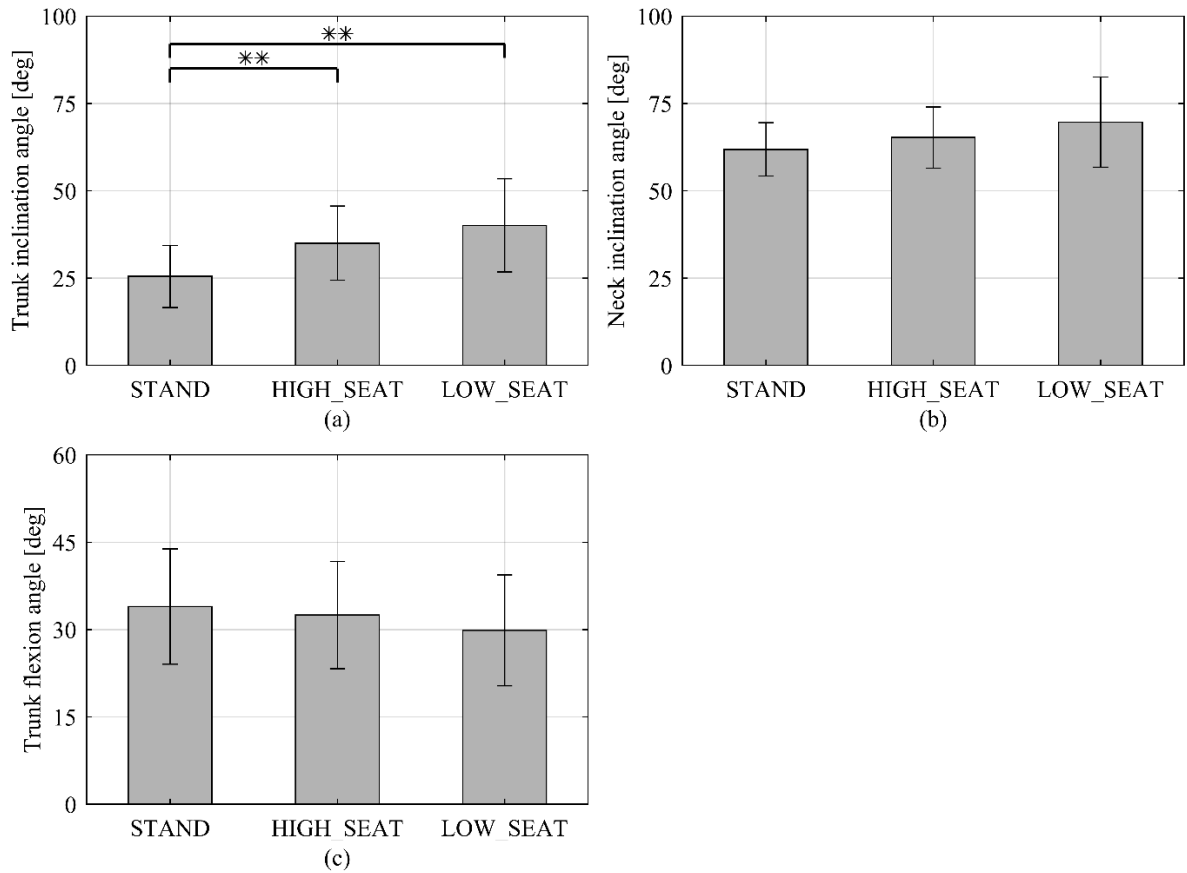

Figure 6. (a) Trunk inclination angle, (b) neck inclination angle, and (c) trunk flexion angle. *STAND* is the condition of standing without the exoskeleton. *HIGH_SEAT* and *LOW_SEAT* are such that the knee joint flexion angles when wearing the exoskeleton are set to 125° and 90°, respectively. The significant differences between the conditions are indicated by * ($p < 0.05$) and ** ($p < 0.01$).

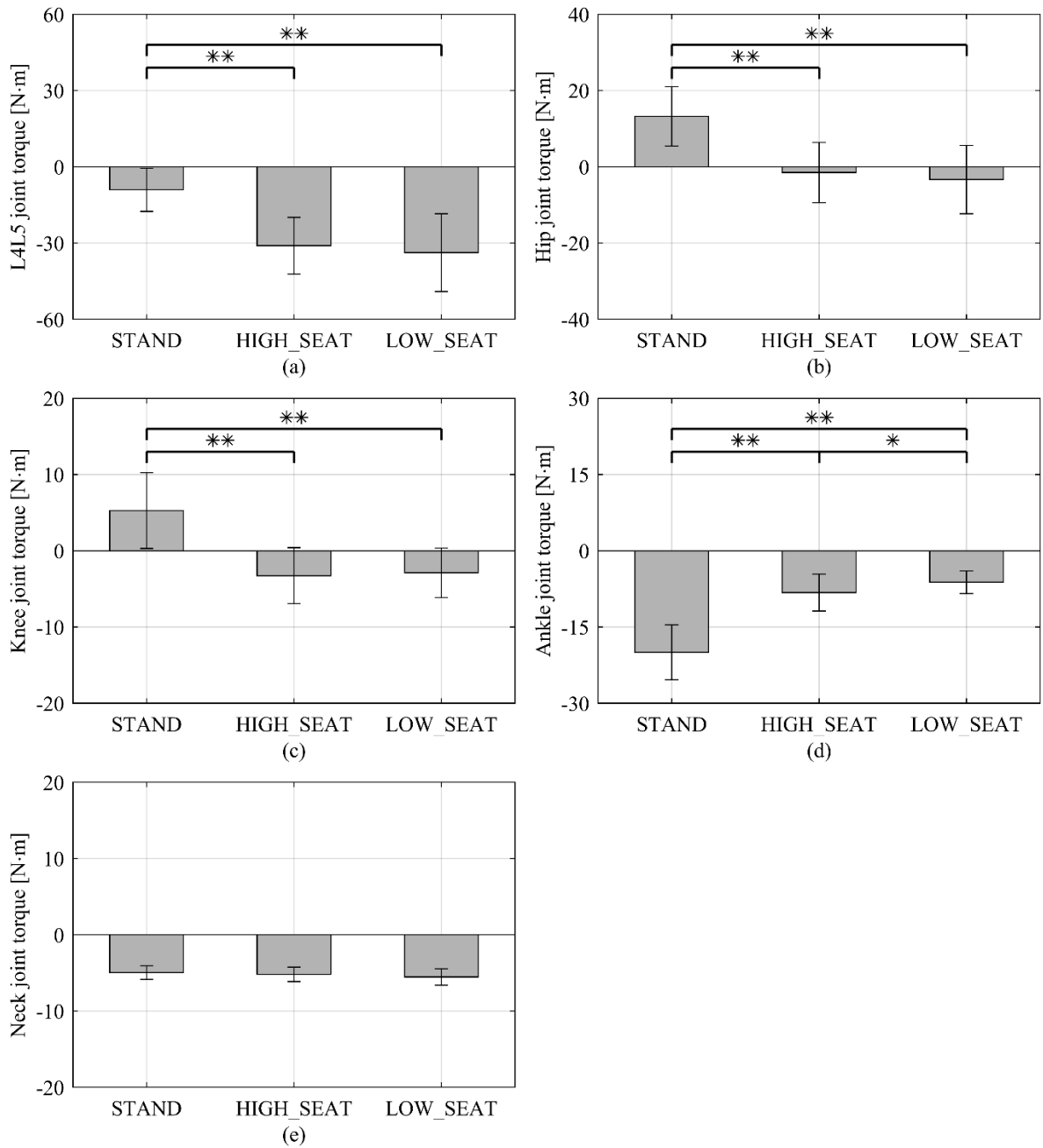

Figure 7. Joint flexion-extension torque of the (a) L4L5, (b) hip, and (c) knee joints. (d) Joint dorsiflexion-plantarflexion torque of the ankle. (e) Joint flexion-extension torque of the neck.

The flexion and dorsiflexion torque are positive. *STAND* is the condition of standing without the exoskeleton. *HIGH_SEAT* and *LOW_SEAT* are such that the knee joint flexion angles when wearing the exoskeleton are set to 125° and 90°, respectively. The significant differences between the conditions are indicated by * ($p < 0.05$) and ** ($p < 0.01$).

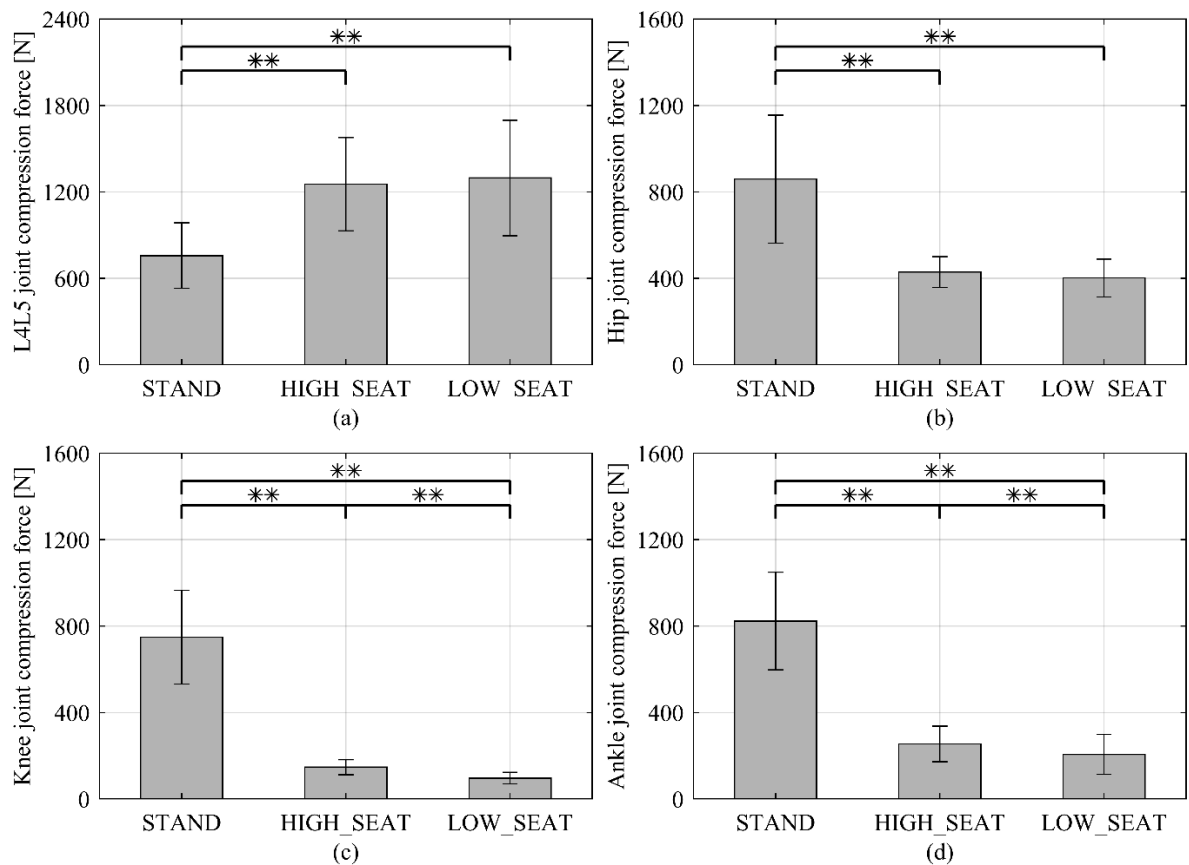

Figure 8. Joint compression force of the (a) L4L5, (b) hip, (c) knee, and (d) ankle joints. *STAND* is the condition of standing without the exoskeleton. *HIGH_SEAT* and *LOW_SEAT* are such that the knee joint flexion angles when wearing the exoskeleton are set to 125° and 90°, respectively. The significant differences between the conditions are indicated by * ($p < 0.05$) and ** ($p < 0.01$).

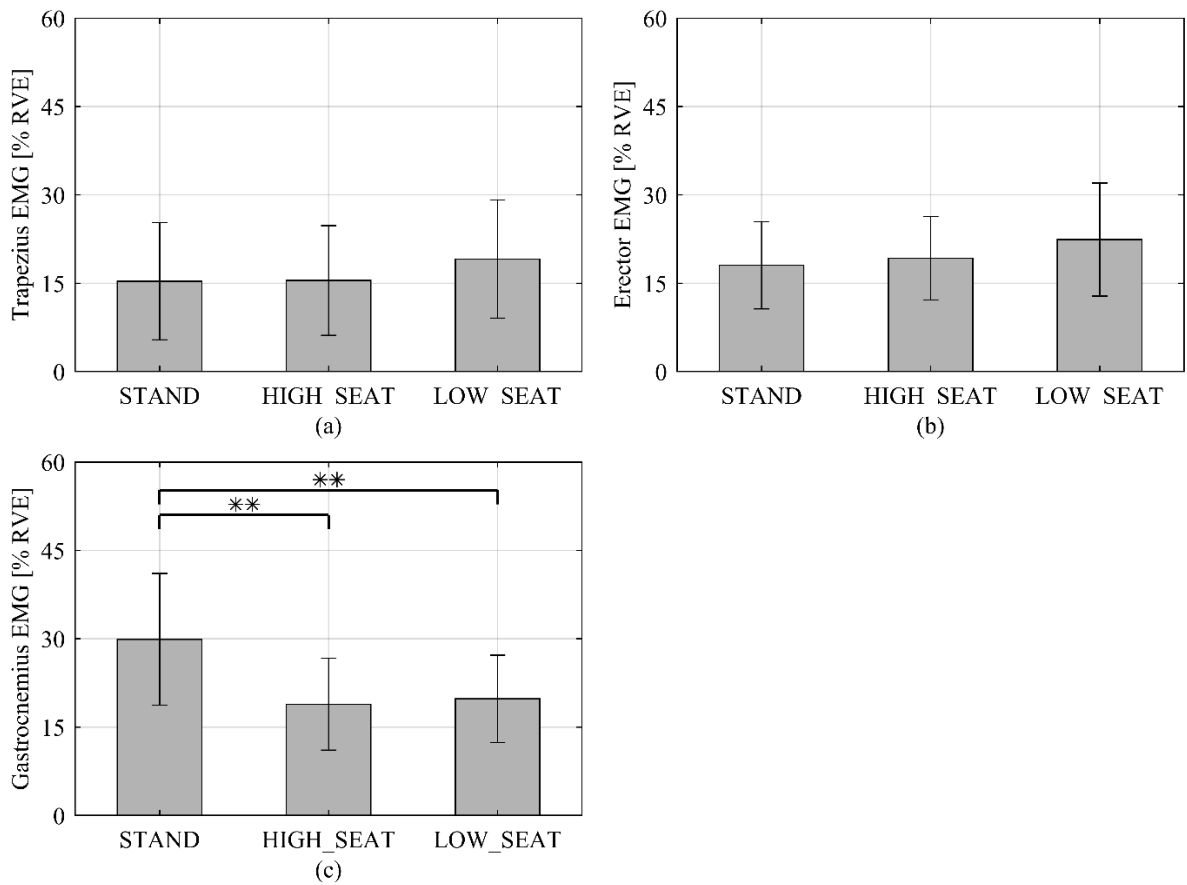

Figure 9. Muscular activity of (a) the upper trapezius, (b) the erector spinae, and (c) the medial head of the gastrocnemius. *STAND* is the condition of standing without the exoskeleton. *HIGH_SEAT* and *LOW_SEAT* are such that the knee joint flexion angles when wearing the exoskeleton are set to 125° and 90°, respectively. The significant differences between the conditions are indicated by * ($p < 0.05$) and ** ($p < 0.01$).

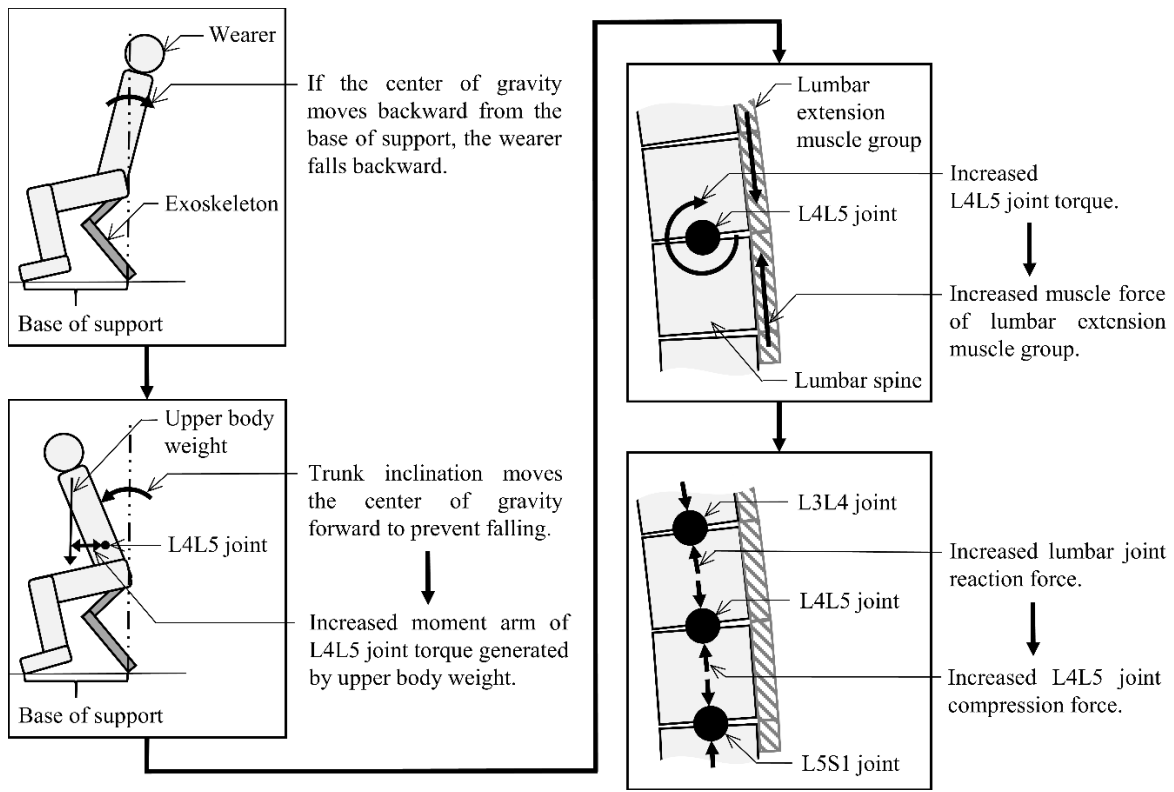

Figure 10.  Overview of the mechanism of the joint compression force increase. The increase in the L4L5 joint torque and the joint compression force is due to the trunk inclination angle, which is critical to maintaining balance when wearing the exoskeleton.